\documentclass[12pt]{iopart}
\expandafter\let\csname equation*\endcsname=\relax
\expandafter\let\csname endequation*\endcsname=\relax

\usepackage[english]{babel}
\usepackage{amsmath}
\usepackage{placeins}
\usepackage{braket}
\usepackage{cleveref}
\usepackage{graphicx}
\usepackage{booktabs}
\usepackage{iopams} 
\usepackage[euler]{textgreek}
\usepackage{wrapfig}
\Crefname{figure}{Figure}{Figures}
\crefname{figure}{figure}{figures}
\begin{document}

\title{Modelling Rabi oscillations for widefield radiofrequency imaging in nitrogen-vacancy centers in diamond}

\author{Simone Magaletti$^1$, Ludovic Mayer$^1$, Jean-François Roch$^2$, Thierry Debuisschert$^1$}
\address{$^1$Thales Research and Technology, 1 Avenue Augustin Fresnel, Palaiseau Cedex, 91767, France

$^2$Université Paris-Saclay, CNRS, ENS Paris-Saclay, CentraleSupelec, LuMIn, Gif-sur-Yvette, 91190, France }

\ead{thierry.debuisschert@thalesgroup.com}
\vspace{10pt}

\begin{abstract}
In this paper we study the dynamics of an ensemble of nitrogen-vacancy centers in diamond when its photoluminescence is detected by means of a widefield imaging system. We develop a seven-level model and use it to simulate the widefield detection of nitrogen-vacancy centers Rabi oscillations. The simulation results are compared with experimental measurements showing a good agreement. In particular, we use the model to explain the asymmetric shape of the detected Rabi oscillations due to an incomplete repolarization of the nitrogen-vacancy center during the pulse sequence implemented for the detection of Rabi oscillations.  
\end{abstract}

\section{Introduction}
The fast spreading of miniaturized RF-based applications requires the development of tools that are able to map the near-field radiofrequency (RF) radiation generated by electromagnetic devices to analyze and monitor their performance. Function and failure analysis \cite{Tiwari2018} as well as material \cite{Greenawald2000,Joffe2013} and device \cite{Weber2012} characterization are some of the several applications that require the measurement of the spatial distribution of a RF field, with both high spatial resolution and high sensitivity. Recently, quantum sensors based on vapor cells \cite{horsley2015}, ultracold atoms \cite{Bohi2010} and negatively charged nitrogen-vacancy centers (NV) in diamond \cite{Dong2018}, have been used to realize the near-field spatial mapping of the RF radiation. Their working principle relies on measuring Rabi oscillations to retrieve the amplitude of the electric or magnetic field exciting the system. Since Rabi frequency is directly related to the amplitude of the electromagnetic field by physical constants known with high accuracy, this technique does not require any calibration procedure \cite{Paulusse2005,Fantom1990}.
Here we discuss the dynamics of an ensemble of NV centers in diamond when a widefield imaging system is used for the detection of Rabi oscillations and we point out the main differences with the detection method using a confocal optical microscopy set-up operated in photon counting regime with an avalanche photodiode (APD). Furthermore, we develop a model that, taking into account the camera gating time, simulates the NV center dynamics and the photoluminescence (PL) acquisition process, resulting in a versatile tool for applications based on widefield imaging.

\section{Widefield detection of nitrogen-vacancy center Rabi oscillations}
\begin{wrapfigure}{r}{0.5\textwidth}
	\begin{center}
		\includegraphics[width=0.5\textwidth]{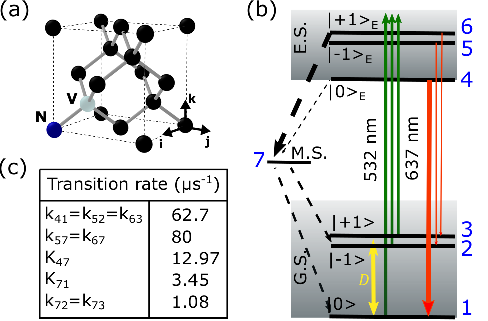}
	\end{center}
	\caption{(a) NV center in the diamond lattice. The blue sphere represents the nitrogen atom, the white sphere the associated carbon vacancy and the black spheres are the carbon atoms. (b) NV center energy levels. (c) NV center transition rates \cite{Robledo2011,Duarte2021}. $k_{ij}$ corresponds to the transition rate from level i to level j, according to the level numbering defined in (b).}
	\label{fig:fig1}
\end{wrapfigure}
The negatively charged nitrogen-vacancy center is a spin-1 diamond defect composed of a nitrogen atom and a carbon vacancy in two adjacent lattice positions \cite{Doherty_2013} (\cref{fig:fig1}a). The nitrogen-vacancy axis (NV axis) defines the intrinsic spin quantification axis in absence of static magnetic field. The negatively charged NV center energy level structure is depicted in \cref{fig:fig1}b. The ground state (G.S.) and the excited state (E.S.) are spin triplets while the metastable state (M.S.) is a spin singlet. Spin states are labelled as $\ket{m_s}$, where $m_s$ is the spin projection quantum number along the NV axis. When no static magnetic field is applied, the ground state energy difference between the $\ket{0}$ and the two degenerate $\ket{\pm1}$ spin sublevels corresponds to a Bohr frequency of 2.87~GHz. All sublevels are equally populated at room temperature. Under optical pumping, two decay paths exist: a radiative spin-conserving decay, which is responsible of the PL, and a non-radiative intersystem crossing through the metastable singlet state. The latter is more probable when the NV center is in $\ket{\pm 1}$ (\cref{fig:fig1}c) and allows both the optical polarization of the NV center in $\ket{0}$ and the optical readout of the spin state by monitoring the PL emission rate of the system \cite{Thiering2018}. The resonance frequencies of the two ground state transitions ($\ket{0}\rightarrow\ket{\pm 1}$) allowed by the spin selection rules ($\Delta m_s=\pm1$) can be controlled by a static magnetic field through the Zeeman effect. Moreover, owing to its long coherence time (T\textsubscript{2} in the millisecond range) at room temperature, the NV center spin can be coherently manipulated by a RF field making possible the optical detection of Rabi oscillations \cite{Hanson2008}.

\begin{figure}
	\centering
	\includegraphics[width=\textwidth]{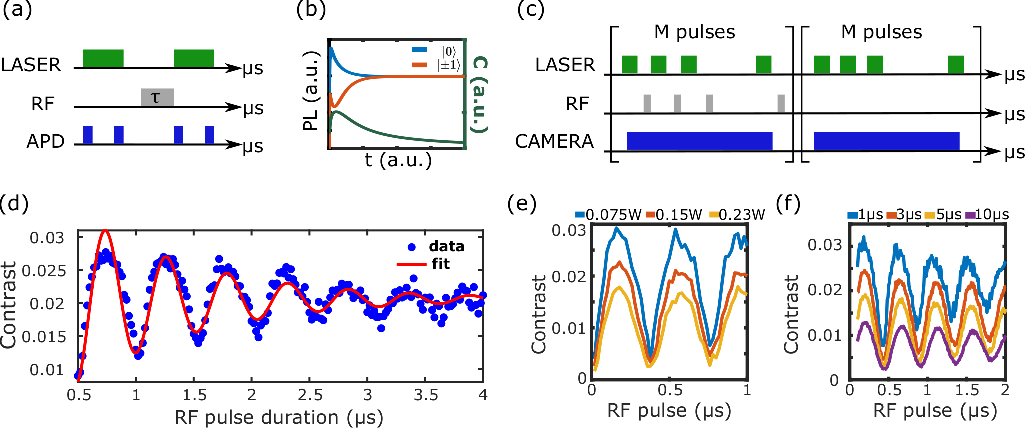}
	\caption{NV center Rabi oscillations. (a) Pulse sequence for the detection of NV centers Rabi oscillations by means of an APD. 
	(b) NV center PL dynamics when the NV center is initially in $\ket{0}$ (blue) or in $\ket{\pm1}$ (orange). In green, time dependence of the related contrast (\cref{eq:Contrast}). (c) Pulse sequence for the detection of NV centers Rabi oscillations by means of a camera. (d) Experimental contrast for different RF pulse durations. The fit is realized using \cref{eq:FitRabi} and we obtain the following parameters:
	$a_\text{R}=0.0206\pm 0.0002$, $b_\text{R}=1.08\pm 0.08$~\textmu s and $c_\text{R}~=~(12.0\pm 0.2)\cdot 10^6$ rad$\cdot$ s\textsuperscript{-1}. $d_\text{R}$ is a phase term which accounts for the fact that, for technical reason, the acquisition started at 0.5~\textmu s instead of 0~\textmu s.
	 (e-f) Experimentally measured contrast as a function of the RF pulse duration, for different laser powers (e) and different laser pulse durations (f). Asymmetric Rabi oscillations are observed for low power or short duration laser pulses.}
	\label{fig:fig2}
\end{figure}

A standard detection scheme for NV center Rabi oscillations is depicted in \cref{fig:fig2}a. A laser pulse with a duration of a few \textmu s polarizes the system in $\ket{0}$. After a waiting time of a few hundreds of ns, necessary for the depletion of the metastable state, the system is excited by a RF pulse of duration $\tau$~on resonance with one of the two NV center ground state spin transitions. The cycle is then repeated sweeping $\tau$. At the beginning of each laser pulse an APD integrates the PL over a duration $T$ ($N$\textsubscript{signal}) in order to readout the NV center spin state and, at the end of the pulse, when the system is polarized in $\ket{0}$, the APD is opened again to take a reference signal ($N$\textsubscript{reference}).
The optically detected magnetic resonance (ODMR) contrast is defined as:
\begin{equation}
C=\frac{N_{\text{reference}}-N_{\text{signal}}}{N_{\text{reference}}}
\label{eq:Contrast}
\end{equation}
Owing to the time dependence of the PL (\cref{fig:fig2}b), the contrast is a function of time. For long integration time, the contrast tends to zero due to the repolarization of the NV centers in $\ket{0}$ \cite{Gupta2016}. The contrast is maximized by optimizing the opening time of the APD, usually some hundreds of ns \cite{Dreau2011}. This measurement scheme was used in Ref.~\cite{Appel2015} to image the RF field generated by a micrometer-scale RF stripline by means of a scanning diamond tip with an NV center at its apex. 

RF field imaging can also be realized using an NV center ensemble in bulk diamond and a widefield imaging system to avoid any scanning procedure, leading to a multiplex advantage in the acquisition time of the RF magnetic field map \cite{Horsley2018}. In this case, an alternative pulse scheme is needed to account for the much larger gating time of a camera compared to an APD (\cref{fig:fig2}c). 
For a single camera exposure, several cycles consisting of a laser pulse, a waiting time for the relaxation of the metastable state and a RF pulse of duration $\tau$~are repeated. The same sequence is then applied with the RF field off to take a reference image. For each RF pulse duration, the ODMR contrast is defined following \cref{eq:Contrast}. 
We realized the experiment using a laser pulse duration of 1~\textmu s, a waiting time of 400 ns and we swept the RF pulse duration from 500 ns to 4~\textmu s (Appendix A). 
The contrast evolution for different values of $\tau$~at a given position in the diamond is plotted in \cref{fig:fig2}d, where we recognize the expected oscillating behavior due to Rabi oscillations. Data are fitted using the function \cite{Appel2015}: 
\begin{equation}
f(\tau)=a_\text{R}\cdot\left(1-e^{-\tau/b_\text{R}}\cdot\cos\left[c_\text{R}\cdot\tau+d_\text{R}\right]\right)
\label{eq:FitRabi}
\end{equation}
where $a$\textsubscript{R}, $b$\textsubscript{R}, $c$\textsubscript{R}, $d$\textsubscript{R} are respectively the amplitude, the decay time, the angular frequency and the phase of the Rabi oscillations \cite{DeRaedt2012}.
\Cref{eq:FitRabi} partially fits data that exhibit an asymmetric shape. We repeated the measurement for different laser powers (\cref{fig:fig2}e) and laser pulse durations (\cref{fig:fig2}f) showing that the increase of these two parameters leads to more symmetric but less contrasted oscillations between the spin states. The decrease in contrast is directly related to the repolarization of the NV centers in $\ket{0}$, as already described in \cref{fig:fig2}b. In fact, differently from the previously described scheme based on APDs, in this case the detector is opened during the entire laser pulse sequence. As a consequence, the spin readout and the NV center repolarization occur simultaneously and cannot be addressed separately. Therefore a compromise has to be found between either a full repolarization of the NV centers or a high ODMR contrast. To have a better understanding of these mechanisms as well as of the asymmetric shape of the Rabi oscillations, we model the dynamics of an ensemble of NV centers when a widefield imaging system is employed for the detection of the Rabi oscillations. 

\section{Simulation}
The pulse sequence implemented in our experiment consists of three steps:
\begin{itemize}
	\item The application of the laser pulse; 
	\item The depletion of the metastable state;
	\item The application of the RF pulse.
\end{itemize}
For the first two steps, the NV center can be modelled as a seven-level system and its dynamics is described by rate equations that neglect the spin coherence terms~\cite{tetienne2012,Dumeige2013}. On the contrary, the last step only addresses one of the two ground state spin transitions ($\ket{0}\rightarrow\ket{-1}$ in our case) and can be described by the Bloch equations for a two level system \cite{cohen}. Merging the two approaches \cite{Jensen2013}, we can write the temporal evolution of the density matrix of the system ($\rho$) as:
\begin{equation}
\label{eq:RhoEvolution}
	\frac{d\rho}{dt}=\frac{1}{\imath\hbar}\left[H,\rho\right]+\left\lbrace\frac{d\rho}{dt}\right\rbrace_\text{{incoh}}  
\end{equation} 
In this equation, $H$ is the rotating-wave approximation of the Hamiltonian describing the interaction between the NV center and a RF field on resonance with the ground state spin transition $\ket{0}\rightarrow\ket{-1}$~\cite{Dreau2011}. It results:
\begin{equation}
	H=\frac{\hbar\Omega_R}{2}\left(\ket{0}\bra{-1}\e^{i\omega_0t}+\ket{-1}\bra{0}\e^{-i\omega_0t}\right)
\end{equation}
where \textomega\textsubscript{0} is the angular frequency of the spin resonance and \textOmega\textsubscript{R} is the Rabi angular frequency.
 The last term $\{\frac{d\rho}{dt}\}_\text{{incoh}}$ accounts for the optical pumping and the relaxation due to interaction with the environment. We then obtain the following set of equations:
\begin{subequations}
	\label{eq:SimulationModel}
	\begin{align}
	\frac{dn_1}{dt}&=-n_1W_p+n_4k_{41}+n_7k_{71}+\Omega_R\cdot n_c\\
	\frac{dn_2}{dt}&=-n_2W_p+n_5k_{52}+n_7k_{72}-\Omega_R\cdot n_c \\
	\frac{dn_3}{dt}&=-n_3W_p+n_6k_{63}+n_7k_{73} \\
	\frac{dn_4}{dt}&=n_1W_p-n_4k_{41}-n_4k_{47} \\
	\frac{dn_5}{dt}&=n_2W_p-n_5k_{52}-n_5k_{57} \\
	\frac{dn_6}{dt}&=n_3W_p-n_6k_{63}-n_6k_{67} \\
	1&=n_1+n_2+n_3+n_4+n_5+n_6+n_7\\
	\frac{dn_c}{dt}&=-\Gamma_2\cdot n_c+\frac{\Omega_R}{2}\cdot(n_2-n_1)
	\end{align}
\end{subequations}
where $n_i=\rho_{ii}$ is the NV center population of the level $i$ (\cref{fig:fig2}b) and $n_c=\Im[\rho_{21}\text{e}^{i\omega_0 t}]$ represents the coherence between level 1 and 2.
$W$\textsubscript{p} is the laser pumping parameter and depends on the laser intensity (Appendix B), and $\Gamma$\textsubscript{2} is the spin decoherence rate. The spin-lattice relaxation, which has a characteristic T\textsubscript{1} time of approximately a few ms, is neglected since it is three order of magnitude longer than the characteristic times of the time evolution in the previous equations. The pump excites a vibronic sideband which decays quickly via phonon emission to levels 4, 5 and 6. This allows us to neglect the down-transition rates due to the pump light. The NV center PL is proportional to the excited state population defined as:
\begin{equation}
n_\text{E}(t)=n_4(t)+n_5(t)+n_6(t)
\label{eq:PL}
\end{equation}

\begin{figure}
	\centering
	\includegraphics[width=0.9\textwidth]{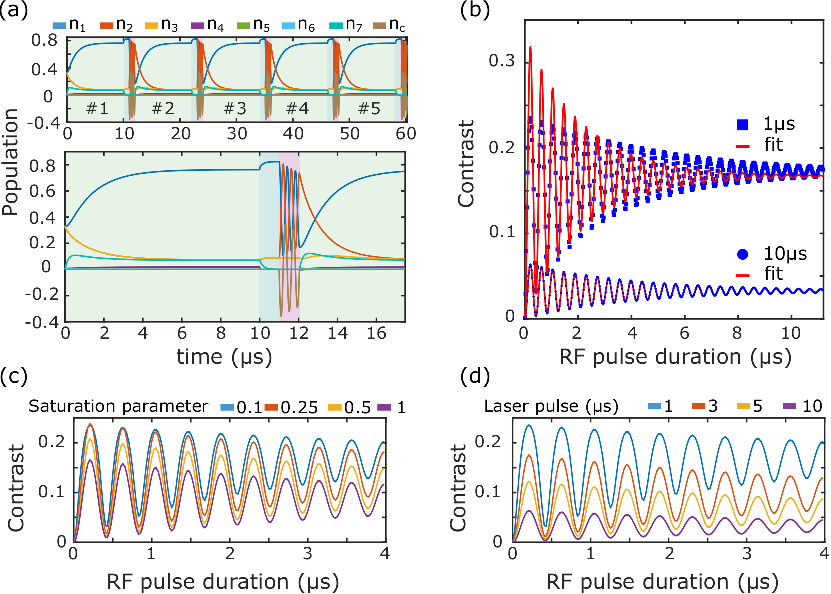}
	\caption{Simulation results. (a) Time evolution of the NV center population. The plot at the bottom is a zoom of the one at the top. Green areas correspond to the laser pulse, light blue areas to the metastable state depletion, pink areas to the RF pulse. $n$\textsubscript{2} and $n$\textsubscript{3} as well as $n$\textsubscript{5} and $n$\textsubscript{6} are often superimposed. (b) Simulation of the detected Rabi oscillations when the laser pulse duration is 10~\textmu s and 1~\textmu s. (c) Rabi oscillations for different saturation parameters and laser pulse duration equal to 1~\textmu s. (d) Rabi oscillations for different laser pulse durations and a saturation parameter equal to 0.1.}
	\label{fig:fig3}
\end{figure}
To account for the several pulse cycles in a single camera exposure, the three steps are iterated several times until a stationary behavior is reached (\cref{fig:fig3}a). For each step, the rate equations are solved using as initial condition the final condition of the previous step. The initial condition of the first step of the first iteration is the thermal equilibrium state at room temperature, that is:
$n~=~[n\textsubscript{1};~n\textsubscript{2};~n\textsubscript{3};~n\textsubscript{4};~n\textsubscript{5};~n\textsubscript{6};~n\textsubscript{7};~n\textsubscript{c}]~=~[1/3;~1/3;~1/3;~0;~0;~0;~0;~0]$.
During the laser pulse we set $\Omega$\textsubscript{R}~=~0, because the RF field is off, and $W$\textsubscript{p}~=~$1.9~\times~10^6$~s\textsuperscript{-1}. This value is chosen according to our experimental condition and corresponds to NV centers pumped at a laser power equal to 10\% of the saturation power (Appendix B). The spin decoherence rate induced by the optical pumping process reads as \cite{Dreau2011}:
\begin{equation}
\Gamma_\text{c}=\Gamma_\text{c}^\infty \frac{s}{s+1}
\end{equation}
where $\Gamma_\text{c}^\infty=8\times10^7$~s\textsuperscript{-1} is the inverse of the excited state lifetime ($\approx$~13~ns); $s$ is the PL saturation parameter defined as the ratio between the laser power used in the experiment and the saturation power. We set $s$ equal to 0.1 (Appendix B) and $\Gamma$\textsubscript{2} equal to $\Gamma$\textsubscript{c}.
During the depletion of the metastable state, the laser and the RF driving field are off ($W$\textsubscript{p}~=~$\Omega$\textsubscript{R}~=~0) and the spin coherence relaxation rate is related to the NV center T\textsubscript{2}\textsuperscript{*} time. We set $Γ_2=\frac{1}{\text{T}_2^*}=5\times10^5~s^{-1}$ assuming T\textsubscript{2}\textsuperscript{*}~=~2~\textmu s, which is the typical value for optical grade diamond \cite{Rosskopf2014}.
Finally, during the RF pulse, we set: $\Omega$\textsubscript{R}~=~1.5$~\times~$10\textsuperscript{7}~rad~$\cdot$s\textsuperscript{-1}, $W$\textsubscript{p}~=~0 and $\Gamma$\textsubscript{2}~=~5$~\times~$10\textsuperscript{5}~s\textsuperscript{-1}. 
The simulation is then performed for different RF pulse durations. The reference signal is obtained by setting $\Omega$\textsubscript{R}~=~0 during all the simulation.

\Cref{fig:fig3}a shows the time evolution of the different level populations when the laser pulse duration is set to 10~\textmu s and both the waiting time for the depletion of the metastable state and the RF pulse duration are set to 1~\textmu s.
During the laser pulse (green areas on \cref{fig:fig3}a) we can observe the polarization of the system in the level 1 as well as an increase of population both in the metastable and in the excited states. For this pumping parameter, the system is polarized after a few \textmu s (Appendix B). Turning off the laser (light blue areas on \cref{fig:fig3}a) we observe the faster decay of the excited states and the slower decay of the metastable state, leading to the increase of the ground states population. Finally, when the RF field is on resonance with the ground state $\ket{0}\rightarrow\ket{-1}$ transition (pink areas on \cref{fig:fig3}a), the populations of these two levels, as well as the spin coherence term n\textsubscript{c} start oscillating. Note that the amplitude of the Rabi oscillations depends on the population difference between $\ket{0}$ and $\ket{-1}$, and is thus directly related to the degree of polarization of the system. Turning the laser on, we restart the pulse sequence. After some iterations, small compared to the total number of iterations during a single exposure of the camera, the system reaches a steady state behaviour. We evaluate the PL intensity (\cref{eq:PL}) and the ODMR contrast (\cref{eq:Contrast}) integrating over the duration of the laser pulse of the last iteration.
\Cref{fig:fig3}b shows the contrast versus the RF pulse duration for two different laser pulse durations: 10~\textmu s, sufficient to fully repolarize the system at the end of the pulse, and 1~\textmu s, too short to completely repolarize the system. The fits are realized using \cref{eq:FitRabi} and we obtain the same Rabi frequency for the two curves, despite the asymmetric shape. The lower contrast for the longest laser pulse is due to the repolarization of the system, as already shown in \cref{fig:fig2}b. 
\Cref{fig:fig3}c-d show the evolution of the ODMR contrast for different pumping parameters and laser pulse durations respectively. In agreement with the experimental data (\cref{fig:fig2}d-f), we observe asymmetric and highly contrasted Rabi oscillations both for short laser pulses and low pumping parameters. 
 
\begin{figure}
	\centering
	\includegraphics[width=\textwidth]{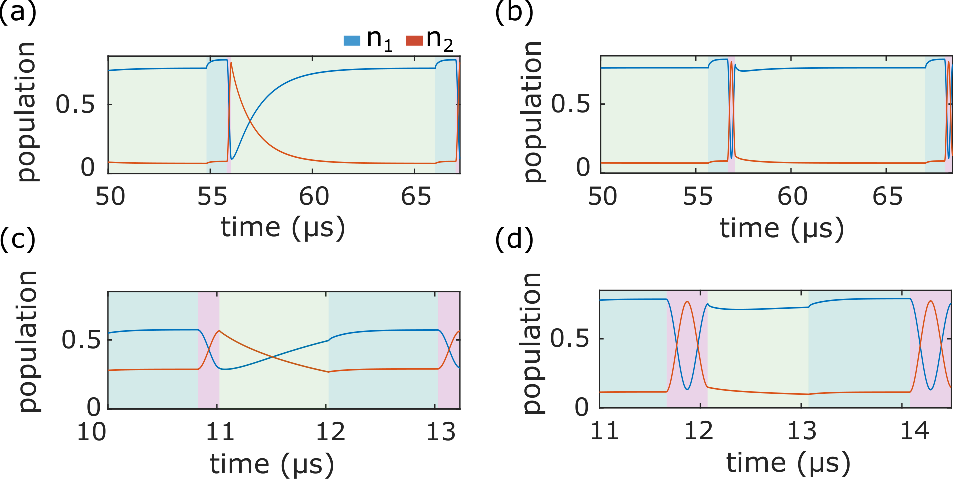}
	\caption{NV center ground state $\ket{0}$ and $\ket{-1}$ population during the last iteration of the simulation, when the system has reached the steady-state. (a) Laser pulse duration 10 \textmu s, RF pulse duration corresponding to a \textpi- pulse. (b) Laser pulse duration 10~\textmu s, RF pulse duration corresponding to a 2\textpi-pulse. (c) Laser pulse duration 1 \textmu s, RF pulse duration corresponding to a \textpi-pulse.(d) Laser pulse duration 1 \textmu s, RF pulse duration corresponding to a 2\textpi-pulse. }
	\label{fig:fig5}
\end{figure}

We further investigate the asymmetry in the Rabi oscillations looking at NV center dynamics when the RF pulse duration corresponds to a \textpi-pulse or a 2\textpi-pulse for two different laser pulse durations: 10~\textmu s (\cref{fig:fig5}a-b) and 1~\textmu s (\cref{fig:fig5}c-d).
A long laser pulse (10~\textmu s) allows a complete repolarization of the NV center after each RF pulse. The polarization is thus independent of the initial state and of the RF pulse itself (\cref{fig:fig5}a-b). A short laser pulse (\cref{fig:fig5}c-d) leads to an incomplete polarization of the NV centers that depends on the RF pulse duration itself. For example, after a 2\textpi-pulse (\cref{fig:fig5}d) the system is easier to fully repolarize than after a  \textpi-pulse (\cref{fig:fig5}c). As a consequence, a short laser pulse may allow a full repolarization in the former case but not in the latter. Since the amplitude of Rabi oscillations (associated to the parameter $a$\textsubscript{R} in \cref{eq:FitRabi}) depends on the polarization of the system before each RF pulse, it is independent of the RF pulse duration $\tau$~when the laser pulse is able to completely repolarize the NV centers; whereas it is a function of $\tau$, causing the asymmetric shape observed in the detected signal (\cref{fig:fig2}d) in the other case. 

\section{Conclusion}
We modelled the widefield detection of NV center Rabi oscillations and used this model to investigate the results obtained in our experiment. Although the time response of the camera is much lower than the Rabi oscillations period, we are nevertheless able to observe those ones using a proper pulse sequence. In particular, we showed that the asymmetric shape of the detected Rabi oscillations is due to an incomplete repolarization of the NV centers at the end of the laser pulse when this latter is too short. More generally, the model allows studying and visualizing the behavior of an ensemble of NV center taking into account the specificities of the widefield detection procedure, namely a collection of the fluorescence over the whole duration of the pump laser pulse and a gating-time of the camera much longer than the characteristic time of the NV centers Rabi oscillations. It results in a powerful and versatile tool to improve widefield Rabi oscillations detection and thus the imaging of RF field. 
Combining the coherent control of the spin with the multiplexing advantages of the widefield imaging opens the way to new applications based on NV centers in diamond.
\appendix
\section{Experimental set-up}
The diamond is an optical grade CVD sample (Element six) with two \{100\} main faces and four \{110\} facets. It is pumped by a 532~nm laser modulated by an AOM in a double pass configuration. Total losses on the pumping branch are approximately 3dB. The PL is collected from a \{110\} facet through an imaging system composed of a microscope objective (20$\times$, 0.33 NA), a polarizer, an optical filter (FF01-697-75 nm band-pass filter, Semrock) to suppress the laser backscattered light and the PL of the neutrally charged nitrogen vacancy center, and a CMOS camera (UI-5240CP-M, IDS). Two neodymium magnets generate an homogeneous static magnetic field in proximity of the NV centers imaged by the camera. The magnetic field is oriented along one of the two NV center families laying on the \{110\} plane. The magnetic field is set approximately at 51 mT, near the excited state level anticrossing, to optically polarize the NV center nitrogen nuclear spin \cite{Jacques2009}. In this condition, the hyperfine structure of the NV center, which results from the interaction between the NV center electron spin and the nitrogen nuclear spin, and which for \textsuperscript{14}N consists in three hyperfine peaks, reduces to a single peak, making the NV center almost equivalent to a two-level system\cite{Dreau2011}. The RF field is generated by a RF generator, pulsed through a RF switch and brought in proximity of the NV centers through a loop antenna. The RF switch and the AOM generator are synchronized by means of a delay generator.

For each RF pulse duration, several images are acquired both with the RF field on and off to improve the signal-to-noise ratio of the measurement.
\begin{figure}
\centering
	\includegraphics[width=0.5\textwidth]{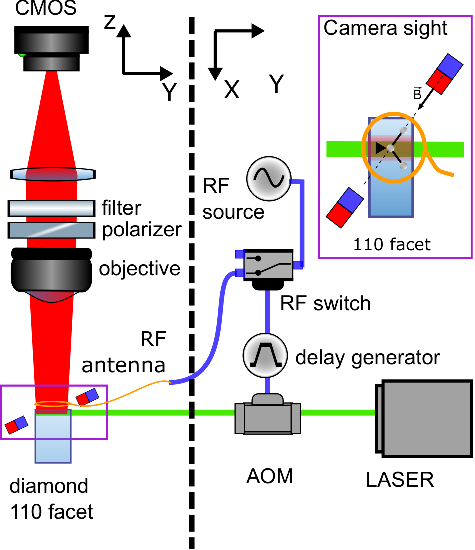}
	\caption{Experimental set-up. On the left of the dashed line the imaging branch of the experimental set-up. On the right of the dashed line the RF and the optical pumping branches. The violet top right inset shows the view from the camera of the diamond area. The magnetic field is aligned along one of the two NV center families laying in the diamond \{110\} plane}
\label{fig:figAppA}
\end{figure}

\section{NV center dynamics}
\label{AppendixModel}
\begin{figure}
	\centering
	\includegraphics[width=\textwidth]{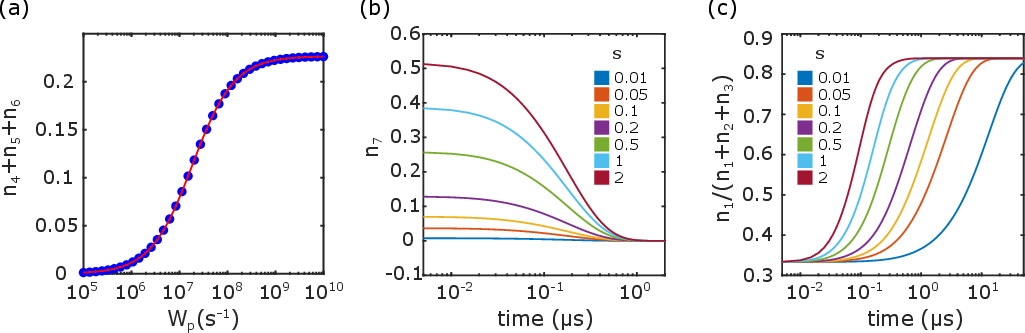}
	\caption{Simulation of the NV centers dynamics. (a) Steady-state solution of NV center excited state for different value of $W$\textsubscript{p} (b) Time evolution of the metastable state population after the laser is turned off for different values of the saturation parameter $s$ (c) Time evolution of the ground state polarization when the laser is turned on for different values of the saturation parameter s.}
	\label{fig:figAppB}
\end{figure}
We solve \cref{eq:SimulationModel} in the stationary state condition for different values of $W$\textsubscript{p}. In \cref{fig:figAppB}a we plot the excited state population ($n$\textsubscript{E}=$n$\textsubscript{4}+$n$\textsubscript{5}+$n$\textsubscript{6}, proportional to the emitted PL) for different values of $W$\textsubscript{p}. We fit the curve by a saturation law: \begin{equation}
n_\text{E}(W_\text{p})=a_\text{p}\cdot\frac{W_\text{p}}{W_\text{p}+W_\text{p}^{\text{Sat}}} 
\end{equation}
where $W$\textsubscript{p}\textsuperscript{Sat} is the value of $W$\textsubscript{p} for which the PL saturates and a\textsubscript{p} is a constant of proportionality. Inserting the typical values of the NV transition rates \cite{Robledo2011,Duarte2021}, we found $W$\textsubscript{p}\textsuperscript{Sat}~=~1.9$~\times~$10\textsuperscript{7} s\textsuperscript{-1}. Thus, we can easily evaluate the saturation intensity ($I$\textsuperscript{Sat}) as: 
\begin{equation}
I^{\text{Sat}}=\frac{W_\text{p}^{\text{Sat}}\cdot c \cdot h}{\sigma\cdot\lambda}
\end{equation}
where $h$ is the Plank constant, $c$ is the speed of light, \textlambda~is the laser wavelength and \textsigma~=~3~$\times$~10\textsuperscript{-21}~m\textsuperscript{2} is the NV center cross section \cite{Wee2007}. We obtain $I$\textsuperscript{Sat}~=~2.3~{mW$\cdot$\textmu m\textsuperscript{-2}}.
Considering our experimental set-up, the laser is focused on the diamond with a beam waist $w_0$~=~18~\textmu m. The saturation power can be calculated using the formula:
\begin{equation}
P^{\text{Sat}}=\frac{\pi w_0^2}{2}\cdot I^{\text{Sat}}
\end{equation}
We obtain $P$\textsuperscript{Sat}~=~1.2~W.
We define the saturation parameter ($s$) as: $s=\frac{P}{P^{\text{Sat}}}$ where $P$ is the laser power used in the experiment, in our case in the range [75; 250]~mW. Therefore $s$ is in the range [0.06; 0.2]. The values of W\textsubscript{P} corresponding to our experiment are thus in the range [10\textsuperscript{6}; 4~$\times$~10\textsuperscript{6}]~s\textsuperscript{-1}.

\Cref{fig:figAppB}b shows the time evolution of the metastable state population when the laser is turned on for different values of the saturation parameter. We can observe that the typical metastable depletion time is around 400~ns, the value used for the experiment.
  
\Cref{fig:figAppB}c shows the time evolution
of the ground state polarization in $\ket{0}$ when the laser is turned on for different value of s. The time needed to completely polarize the system increases reducing the laser power. For a saturation parameter $s$~=~0.1 it is approximately 10 \textmu s. That explains our choice of using a laser pulse duration of 10 \textmu s for our simulations.

\section{Map of the RF near field of a loop antenna}
\label{AppendixAntennaMeasurement}
\begin{figure}
	\centering
	\includegraphics[width=0.8\textwidth]{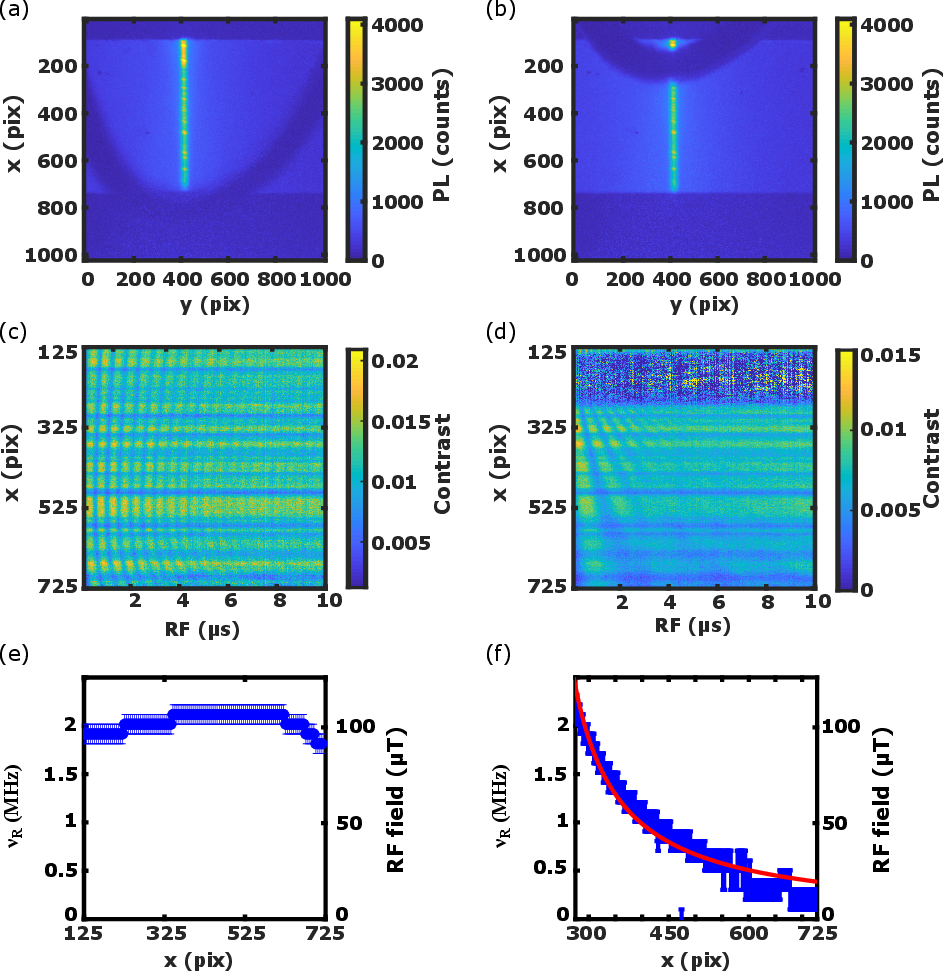}
	\caption{Widefield imaging of the RF field emitted by a loop antenna in the near field region. (a-b) Antenna position. The picture is the full-frame image acquired by the camera. The laser beam (yellow beam in the picture), the shadow of the antenna (dark blue semi-circle) and the diamond (light blue rectangle) are visible. (c-d) Rabi oscillations contrast along the beam propagation axis (x) for the first (c) and second (d) antenna configuration. The large horizontal line showing an almost homogeneous contrast at the top of (d) is the shadow of the antenna. The several thin horizontal lines with low-contrast are caused by inhomogeneities in the NV center concentration \cite{Pham2012}. (e-f) Rabi frequency (\textnu\textsubscript{R}) along the beam propagation axis for the first (e) and the second (f) antenna position. The RF field (B\textsubscript{R}) is calculated using the formula:
		$ B_\text{R}=\frac{\sqrt{2}\nu_\text{R}}{\gamma}$ \cite{Appel2015}, where $\gamma=28$~MHz~$\cdot$~mT\textsuperscript{-1} is the NV center gyromagnetic ratio. The fit (red line) in (f) is realized using \cref{eq:FitFunc}.}
	\label{fig:figAppD}
\end{figure}
We mapped the RF signal emitted in near field by the loop antenna implemented in our set-up (\cref{fig:figAppA}) along the laser beam propagation axis. We investigated two configurations. In the first one, NV centers are excited by the RF field generated inside the wire loop (\cref{fig:figAppD}a) and, in the second one, by the one generated outside of it (\cref{fig:figAppD}b). Rabi oscillations are driven and detected using the pulse sequence in \cref{fig:fig2}c. The laser pulse duration is set to 2~\textmu s in the first configuration and 5~\textmu s in the second one. The RF pulse duration is swept from 0.1~\textmu s to 10~\textmu s in steps of 20 ns.
\Cref{fig:figAppD}c-d show the contrast of Rabi oscillations along the beam propagation axis (x). 
For each x position, the contrast is averaged along y over the 10 pixels around the center of the laser beam. 
We observe that the Rabi frequency depends on the position and it is the sign that the antenna is emitting an inhomogeneous RF field. For each position, the Rabi frequency is evaluated by means of a fast Fourier transform and then plotted in \cref{fig:fig2}e-f. 

Inside the antenna (\cref{fig:fig2}e), the Rabi frequency, and thus the amplitude of the RF field, is almost homogeneous. Outside the antenna, (\cref{fig:fig2}f) the amplitude of the magnetic field decreases while increasing the distance from the wire. The fit function employed in \cref{fig:fig2}f is: \begin{equation}
\label{eq:FitFunc}
f(x)=\frac{a_\text{W}}{x+b_\text{W}-c_\text{W}}
\end{equation} where the parameter $a$\textsubscript{W} accounts for the magnetic field amplitude, $b$\textsubscript{W} is an offset radius due to the mismatch between the zero of the x axis and the center of the wire, and $c$\textsubscript{W} is the origin of the fit set at pixel number 275. We obtain $b$\textsubscript{W}~=~57$\pm$3~\textmu m which is comparable to the radius of the wire (50~\textmu m).

\section*{Acknowledgement}
We thank Nimba Oshnik and Elke Neu for their support in conceiving the experimental set-up. This project received funding from the European Union’s Horizon 2020 research and innovation program under grant agreement No. 820394 (ASTERIQS), the European Union’s Horizon Europe research and innovation program under grant agreement No. 101080136 (AMADEUS), the Marie Skłodowska-Curie grant agreement No. 765267 (QuSCo), the QUANTERA grant agreement ANR-18-QUAN-0008 (MICROSENS), the EIC Pathfinder 2021 program under grant agreement No. 101046911 (QuMicro) and the EMPIR project 20IND05 (QADeT). We acknowledge the support of the DGA/AID under grant agreement ANR- 17-ASTR-0020 (ASPEN).

\section*{Author contributions}
S.M., L.M., J.-F.R., and T.D. contributed to the design and implementation of the research, to the simulations and the analysis of the results and to the writing of the manuscript.

\section*{Conflict of Interest}
The authors declare no conflict of interest.

\section*{Data Availability Statement}
The data and the code that support the findings of this study are available from the corresponding author upon reasonable request.

\section*{Reference}
\bibliographystyle{unsrt}
\bibliography{paperv3} 
\end{document}